\documentclass{article}
\usepackage[margin=2.5cm]{geometry}

\usepackage{rotating}
\usepackage{graphicx}
\usepackage{hyperref}

\newboolean{articletitles}
\setboolean{articletitles}{true} 
\newboolean{uprightparticles}
\setboolean{uprightparticles}{false} 
\newboolean{inbibliography}
\setboolean{inbibliography}{false} 

\begin{document}

\title{\bf LHCb CP violation}

\author{Mika Vesterinen, on behalf of the LHCb collaboration\\ Physikalisches Institut Heidelberg \\ \normalsize Proceedings of the Third Annual Large Hadron Collider Physics Conference LHCP15,\\ \normalsize Aug. 31-Sept 5, 2015 St. Petersburg, Russia.}

\maketitle

\begin{abstract}
The study of $CP$ violation in the beauty hadron sector is a promising
approach to search for the effects of physics beyond the Standard Model.
Several recent measurements in this area from the LHCb experiment are reported in these proceedings.
These are based on the Run-I dataset of 3~fb$^{-1}$ of data collected
at proton-proton centre of mass energies of 7 and 8~TeV.
\end{abstract}

\section{Introduction}
In the Standard Model (SM), the quark flavour transitions are governed
by the $3\times3$ unitary Cabibbo-Kobayashi-Maskawa matrix, 
\begin{equation}
  V_{\rm CKM} =
  \left(
  \begin{array}{ccc}
    V_{ud} & V_{us} & V_{ub} \\
    V_{cd} & V_{cs} & V_{cb} \\
    V_{td} & V_{ts} & V_{tb} \\
  \end{array}\right),
\end{equation}
which relates the mass and flavour eigenstates of the quarks.
The single irreducible phase of this matrix is the source of all $CP$-violation in the quark
sector in the SM.
Six of the nine unitarity constraints form triangles in the complex plane,
one of which is referred to as the {\em unitarity triangle}.
A main goal of the LHCb experiment is to make precision measurements of $CP$-violation,
and to measure the sides and internal angles of the unitarity triangle,
in the hope to reveal anomalies due to the indirect effects of physics beyond the SM.
Several recent LHCb measurements relating to $CP$-violation are reported.
These are based on the Run-I dataset that was recorded during 2011 and 2012
at $pp$ centre of mass energies of $\sqrt{s} = 7$ and $8$~TeV, 
corresponding to integrated luminosities of 1 and 2~fb$^{-1}$, respectively.
In particular, several measurements related to $B_d^0-\bar{B}_d^0$ mixing are reported.
This process occurs through a loop diagram in the Standard Model, and is therefore highly sensitive to the effects of new particles in scenarios beyond the SM.
Throughout these proceedings, charge conjugation is implied unless explicitly stated otherwise.

\section{Measurement of \boldmath{$\sin2\beta$} using \boldmath{$B_d^0 \to  J/\psi K_s^0$}}
\label{Sec:sin2beta}

\begin{figure}[tb]\centering
  \includegraphics[width=0.49\textwidth]{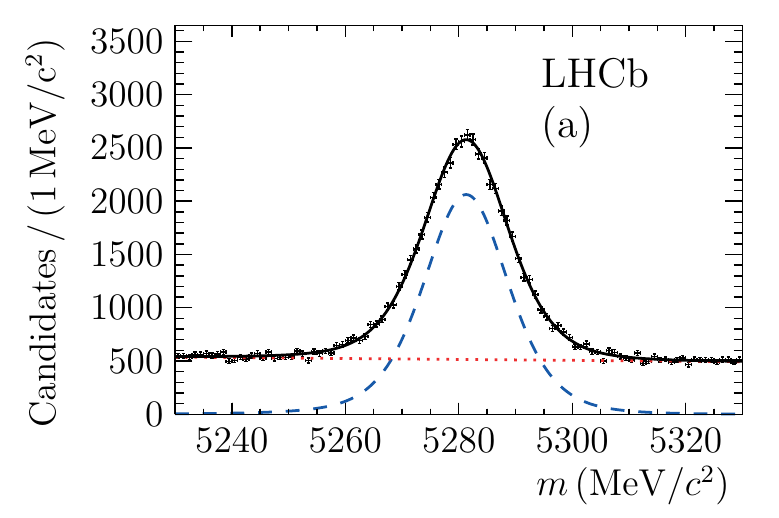}
  \includegraphics[width=0.49\textwidth]{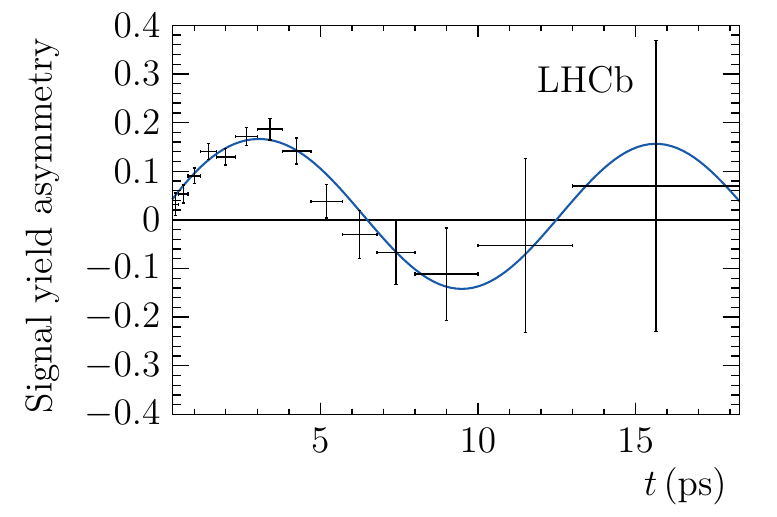}
  
  \caption{\label{fig:sin2beta}Left: the invariant mass spectrum of flavour 
tagged $B_d^0 \to J/\psi K_s^0$ decays~\cite{LHCb-PAPER-2015-034}.
Right: the $CP$ asymmetry of these candidates, after background subtraction, as a function of decay time.}
\end{figure}

The $J/\psi K_s^0$ final state is accessible to both the $B_d^0$ and $\bar{B}_d^0$ mesons.
Allowing for $B_d^0-\bar{B}_d^0$ mixing, there are two amplitudes 
that contribute to this transition, 
and they differ by a relative phase
$\beta = \arg(V_{tb}V^*_{td}/V_{cb}V^*_{cs})$.
The interference of the two amplitudes leads to a 
decay time, $t$, dependent $CP$-asymmetry that oscillates as
$\sin2\beta\cos(\Delta M_d t)$, where $\Delta M_d$ is the $B_d^0-\bar{B}_d^0$ mixing frequency.
LHCb recently reported a measurement of $\sin2\beta$ with the full Run-I
dataset~\cite{LHCB-PAPER-2015-004}, which improves on an earlier measurement
with the 2011 data alone~\cite{Aaij:2012ke}.
A key challenge in this measurement is in the determination of the $B_d^0$ flavour at production.
The new analysis achieves an improved sensitivity by using a new flavour tagging algorithm
that exploits the correlated production of charged pions through the fragmentation process~\cite{Gronau:1992ke,Aaij:2015qla}.
The invariant mass distribution of the flavour tagged $B_d^0 \to J/\psi K_s^0$ candidates
is shown in Figure~\ref{fig:sin2beta} (left), from which a yield of roughly 40k signal decays is extracted.
Figure~\ref{fig:sin2beta} (right) shows, as a function of the decay time, the asymmetry between candidates that are tagged as $\bar{B}_d^0$ at production versus those that are tagged as $B_d^0$. 
An unbinned maximum likelihood fit to the decay time distribution that incorporates information
about the flavour tagging yields the result,
\begin{equation}
\sin2\beta = 0.731 \pm 0.035 \pm 0.020,
\end{equation}
whose precision is comparable to that of the two best measurements from 
the BaBar and Belle experiments~\cite{Aubert:2009aw,PhysRevLett.108.171802}.
The fit also allows for an asymmetry with a cosine dependence on the decay time.
The amplitude of this term is determined to be $C = -0.038 \pm 0.032 \pm 0.005$,
which is compatible with zero as expected in the SM.

\section{Measurement of \boldmath{$\Delta M_d$} using \boldmath{$B_d^0 \to  D^{(*)}\mu\nu X$}}
\label{Sec:DmD}

\begin{figure}[tb]\centering
  \includegraphics[width=0.41\textwidth]{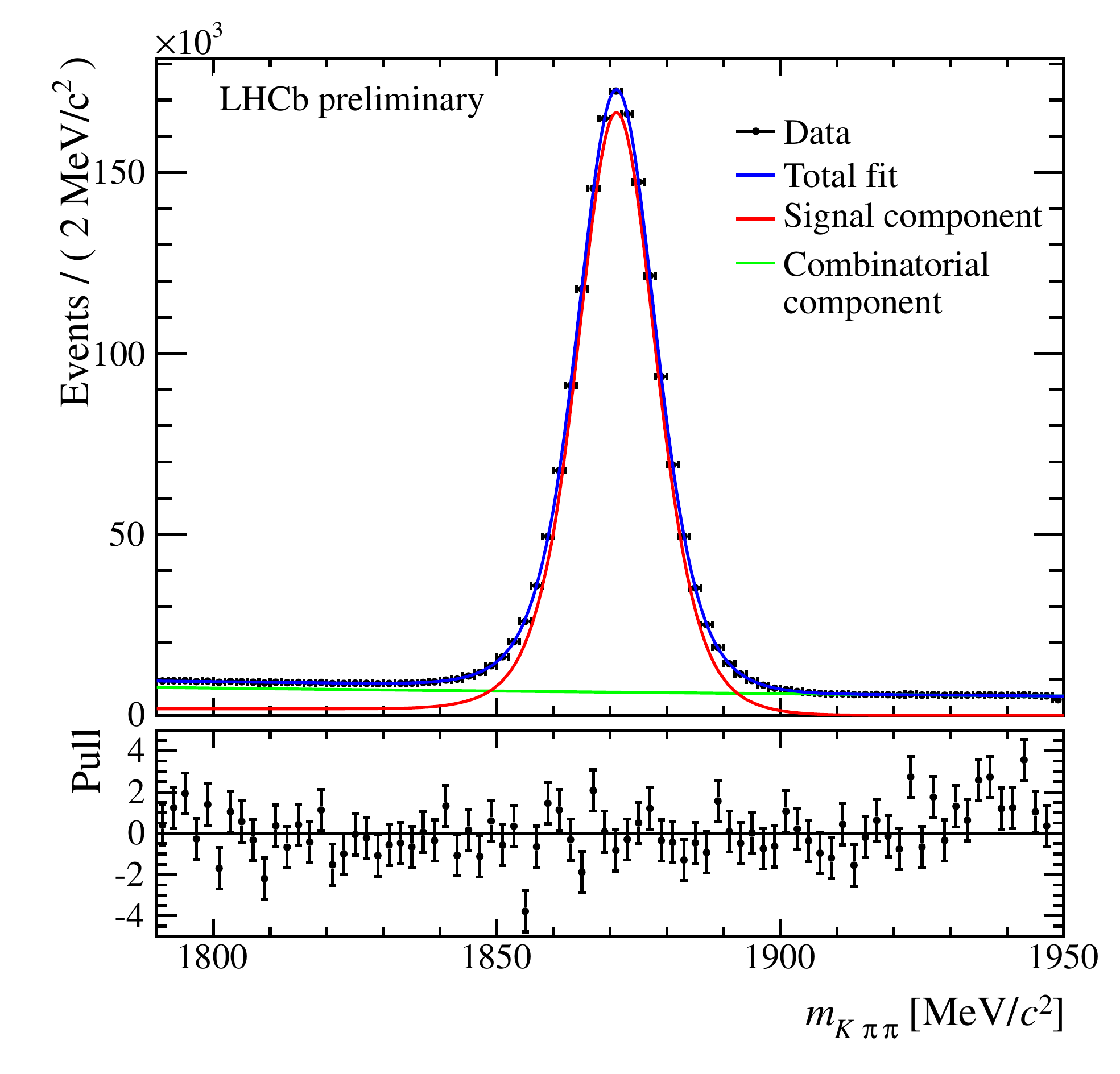}
  \includegraphics[width=0.57\textwidth]{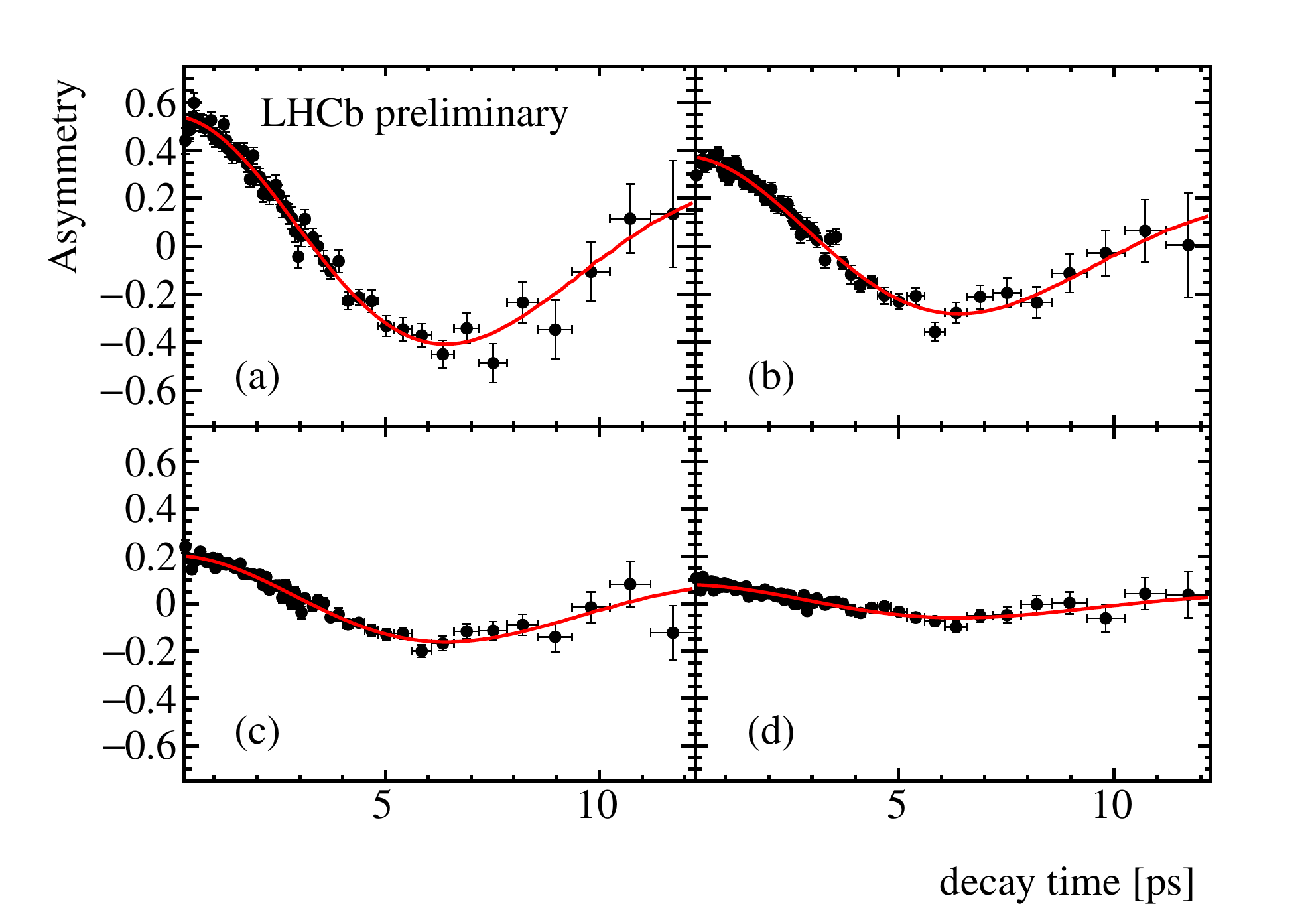}
  
  \caption{\label{fig:DmD}Left: the $D^+$ invariant mass spectrum in
candidate semileptonic $B_d^0$ meson decays~\cite{LHCb-CONF-2015-003}. Right: The mixing asymmetry
as defined in the text.}
\end{figure}

The semileptonic decays $B_d^0 \to D^{(*)-}\mu^+\nu_{\mu} X$ are flavour specific, meaning that the charges of the final state particles unambiguously tag the parent particle as being a $B_d^0$ or a $\bar{B}_d^0$.
They are well suited to the measurement of the
$B_d^0-\bar{B}_d^0$ mixing frequency, $\Delta M_d$.
Using the flavour tagging algorithms described in the previous Section,
these decays can be tagged as either mixed or unmixed.
The mixing asymmetry is defined as,
\begin{equation}
\frac{N_{\rm unmixed}(t) - N_{\rm mixed}(t)}{N_{\rm unmixed}(t) + N_{\rm mixed}(t)} = \cos(\Delta M_d t).
\end{equation}
The semileptonic decays provide a large signal of several millions of events
which permits a high statistical precision.
Figure~\ref{fig:DmD} (left) shows the $D^+ \rightarrow K^-\pi^+\pi^+$ invariant
mass distribution in $\bar{B}_d^0 \to D^+\mu^-\bar{\nu}$ candidates.
On the other hand, they are challenging to analyse, since they can only be partially reconstructed due the presence of a neutrino in the final state.
A new measurement of $\Delta M_d$ based on the full Run-I dataset is reported
in~\cite{LHCb-CONF-2015-003}.
The decay time is corrected for the missing momentum using a parameterisation
determined from simulated signal decays.
The background from $B^+$ mesons decaying into the same final state is suppressed
to a level of a few percent by means of a dedicated multivariate classifier that is based on isolation and 
kinematic information.
The mixing asymmetry for a subset of the data is shown in Figure~\ref{fig:DmD} (right).
The four subfigures correspond to different expected mis-tag rates from the 
flavour tagging algorithms. 
The oscillation has the largest amplitude in the top left subfigure which
corresponds to the lowest mis-tag category.
The measurement yields a value of,
\begin{equation}
\Delta M_d = 503.6 \pm 2.0{\rm (stat)} \pm 1.3{\rm (syst)}~\mathrm{ns}^{-1},
\end{equation}
which is the single most precise measurement of this parameter to date.

\section{Penguin study with \boldmath{$B_s \to J/\psi K^*$} decays}
\label{Sec:Penguin}

\begin{figure}[tb]\centering
  \includegraphics[width=0.49\textwidth]{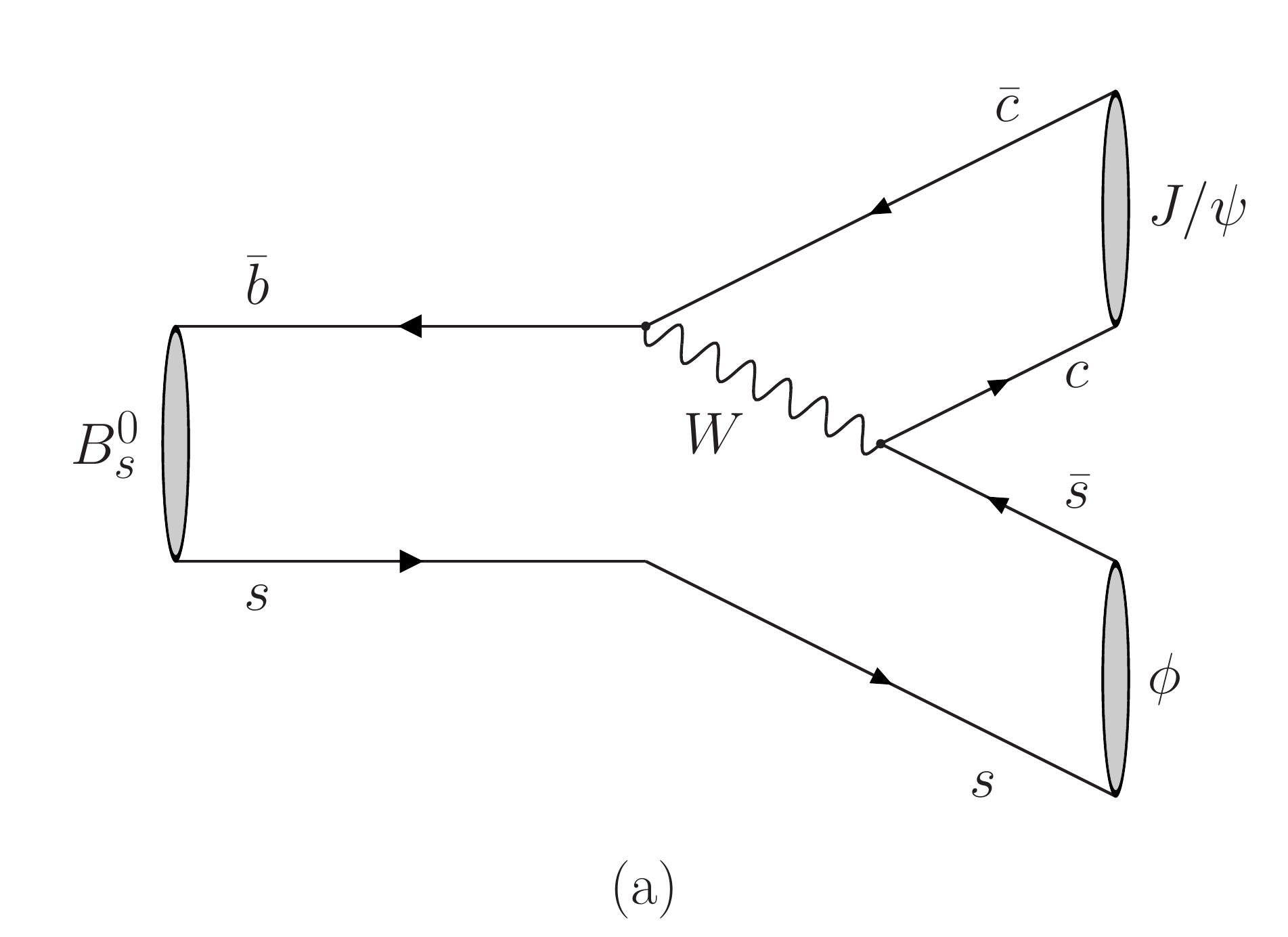}
  \includegraphics[width=0.49\textwidth]{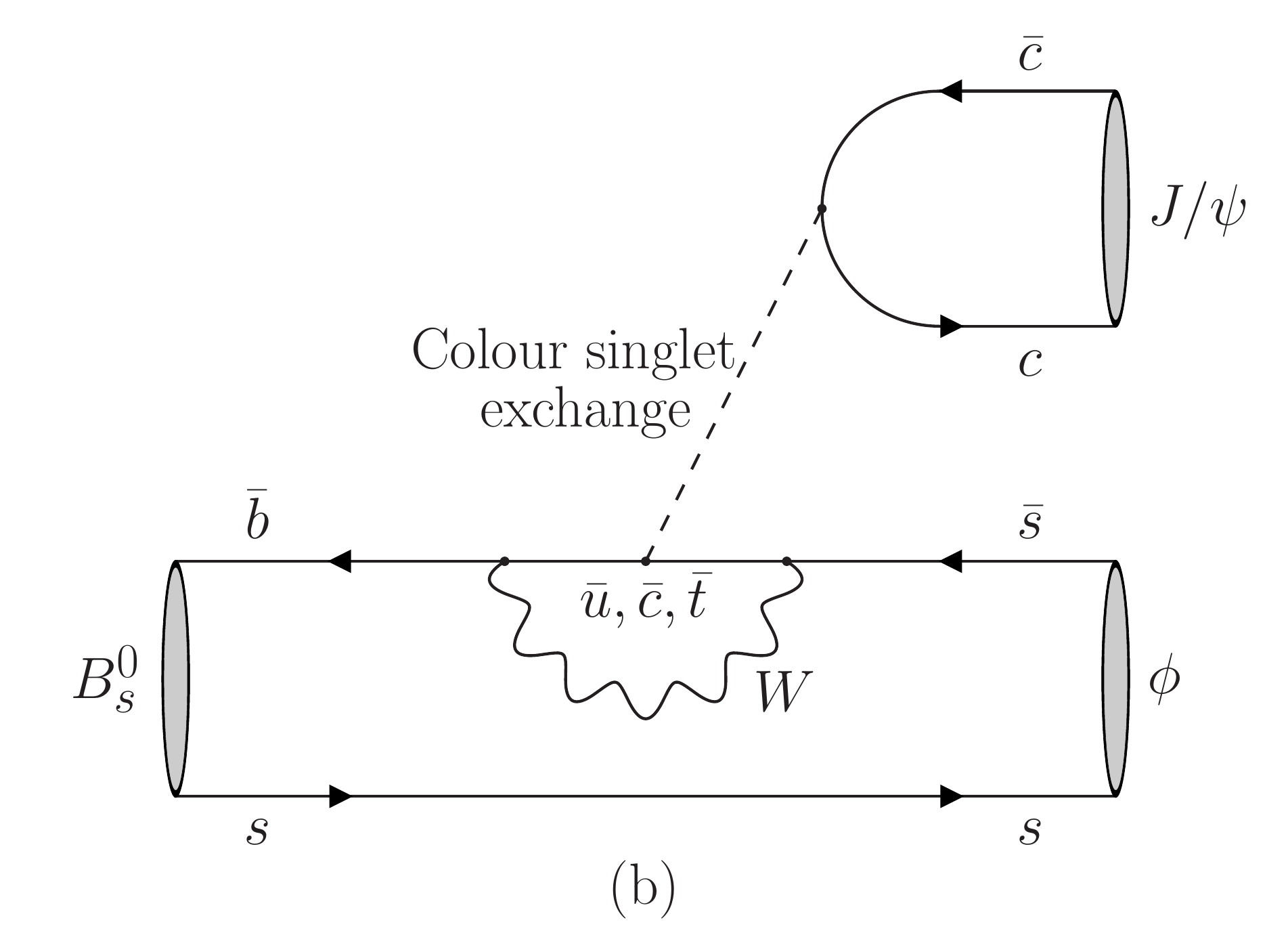}
  
  \caption{\label{fig:Penguin}Left: The leading tree diagram that contributes to the decay
$B_s^0 \to J/\psi \phi$. Right: A penguin diagram contributing to the same decay mode.}
\end{figure}

\begin{figure}[tb]\centering
  \includegraphics[width=0.49\textwidth]{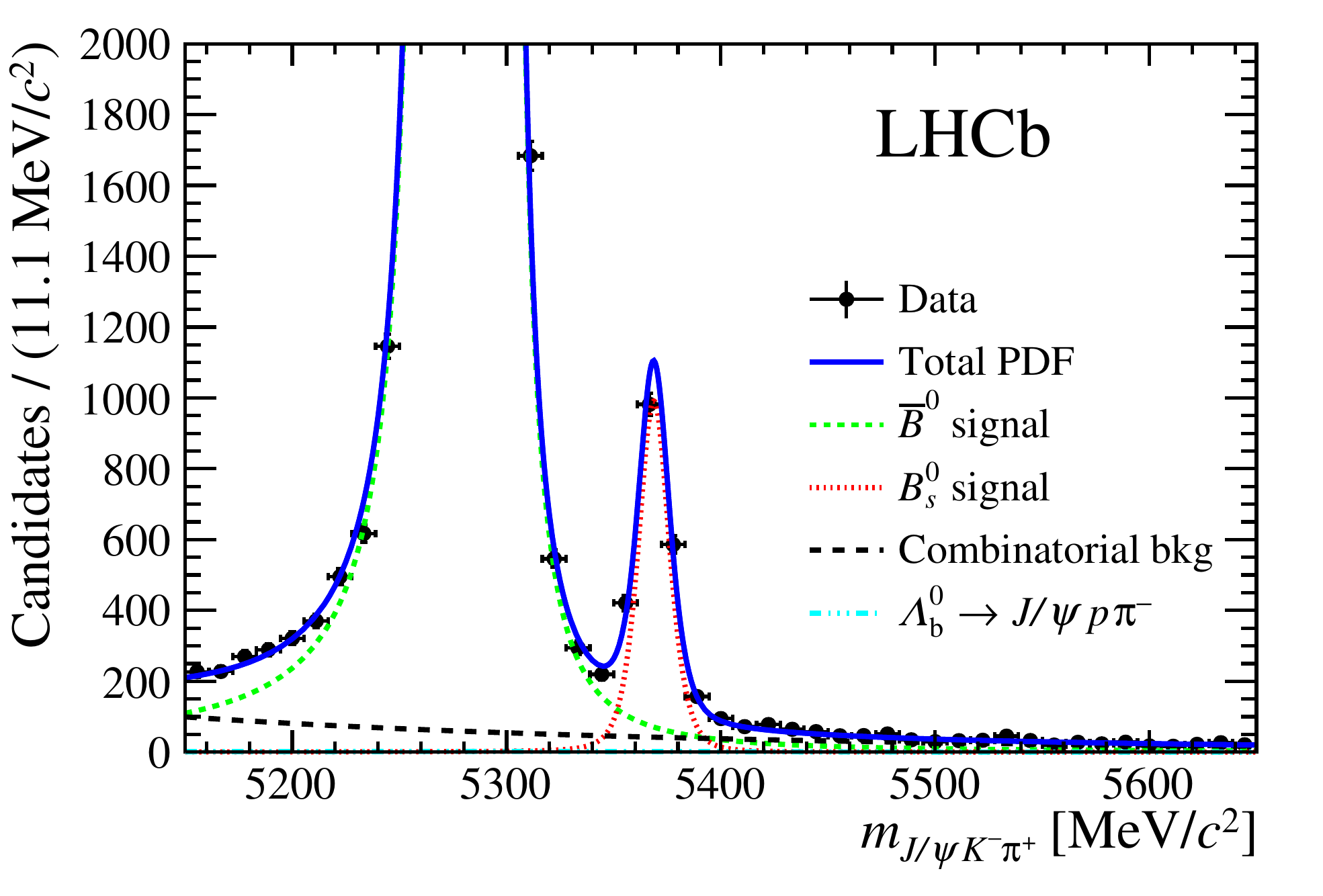}
  
  \caption{\label{fig:JpsiKst}The invariant mass spectrum of $B_s^0 \to J/\psi K^*$ candidates.
The $B_s^0$ signal peak (red) is dwarfed by the larger $B_d^0$ component.}
\end{figure}

A primary goal of LHCb is to precisely measure the phase $\phi_s$,
primarily through the decay $B_s \to J/\psi\phi$.
In the SM, and assuming dominance of the leading tree level diagram
shown in Figure~\ref{fig:Penguin} (left),
$\phi_s = -2\beta_s$, where $\beta_s = \arg(-V_{ts}V_{tb}^*/V_{cs}V_{cb}^*)$.
It is hoped that a precision measurement of $\phi_s$ will reveal an anomaly due to the effects of amplitudes involving particles beyond those of the SM.
There are however further SM contributions, notably from penguin diagrams such as the one shown in Figure~\ref{fig:Penguin} (right).
This ``penguin pollution'' is challenging to calculate, but can be constrained
by studying the related decay $B_s \to J/\psi K^*$ in which the tree diagram
is Cabibbo suppressed.
LHCb recently performed a study of this decay using the full Run-I dataset~\cite{LHCb-PAPER-2015-034}.
Figure~\ref{fig:JpsiKst} shows the invariant mass spectrum of candidate
decays of this type. A clear $B_s$ signal peak can be 
seen on top of the tail from the larger $B_d \to J/\psi K^*$ peak.
The branching fraction of the $B_s$ decay is measured to be,
\[ \mathcal{B}(B_s \to J/\psi \bar{K}^{*0}) = \left( 4.17 \pm 0.18{\rm (stat)} \pm 0.26{\rm (syst)} \pm 0.24{\rm (prod)}\right)\times 10^{-5}, \]
where the last uncertainty relates to the production ratio $f_s/f_d$.
The polarisation fractions are measured to be
\[f_0 = 0.179 \pm 0.027{\rm (stat)} \pm 0.013{\rm (syst)},\]
\[f_{||} = 0.497 \pm 0.027{\rm (stat)} \pm 0.025{\rm (syst)},\]
and the $CP$ asymmetries are measured to be
\[ A^{CP}_0(B_s \to J/\psi \bar{K}^{*0})    = -0.048 \pm 0.057{\rm (stat)} \pm 0.020{\rm (syst)},\]
\[ A^{CP}_{||}(B_s \to J/\psi \bar{K}^{*0}) =  0.171 \pm 0.152{\rm (stat)} \pm 0.028{\rm (syst)},\]
\[ A^{CP}_{\perp}(B_s \to J/\psi \bar{K}^{*0}) =  -0.049 \pm 0.096{\rm (stat)} \pm 0.025{\rm (syst)}.\]
The results of this analysis are combined with those of an earlier LHCb study of the 
$SU(3)$ related decay $B_d^0 \to J/\psi \pi^+\pi^-$~\cite{Aaij:2014vda},
yielding tight constraints on the penguin pollution to $\phi_s$ measured in $B_s \to J/\psi \phi$,
\[ \Delta\phi_{s,0}^{J/\psi\phi} = 0.000^{+0.009}_{-0.011}{\rm (stat)} ^{+0.004}_{-0.009}{\rm (syst)}~\mathrm{rad},\]
\[ \Delta\phi_{s,||}^{J/\psi\phi} = 0.001^{+0.010}_{-0.014}{\rm (stat)} \pm 0.008{\rm (syst)}~\mathrm{rad},\]
\[ \Delta\phi_{s,\perp}^{J/\psi\phi} = 0.003^{+0.010}_{-0.014}{\rm (stat)} \pm 0.008{\rm (syst)}~\mathrm{rad}. \]

\section{Measurement of \boldmath{$\gamma$} with \boldmath{$B^+ \to D h^+ h^- h^+$} decays}
\label{Sec:D3h}

\begin{figure}[tb]\centering
  \includegraphics[width=0.40\textwidth]{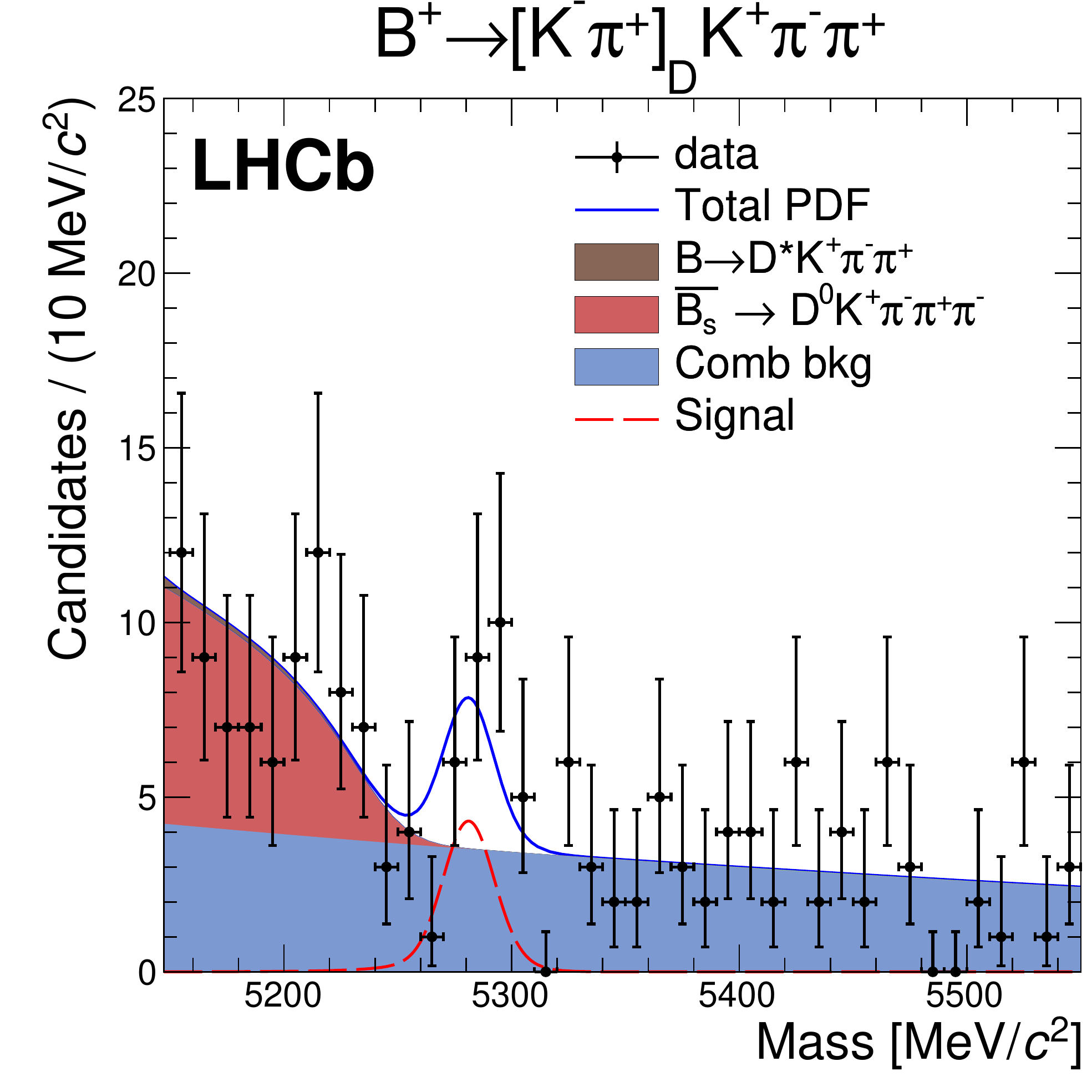}
  \includegraphics[width=0.56\textwidth]{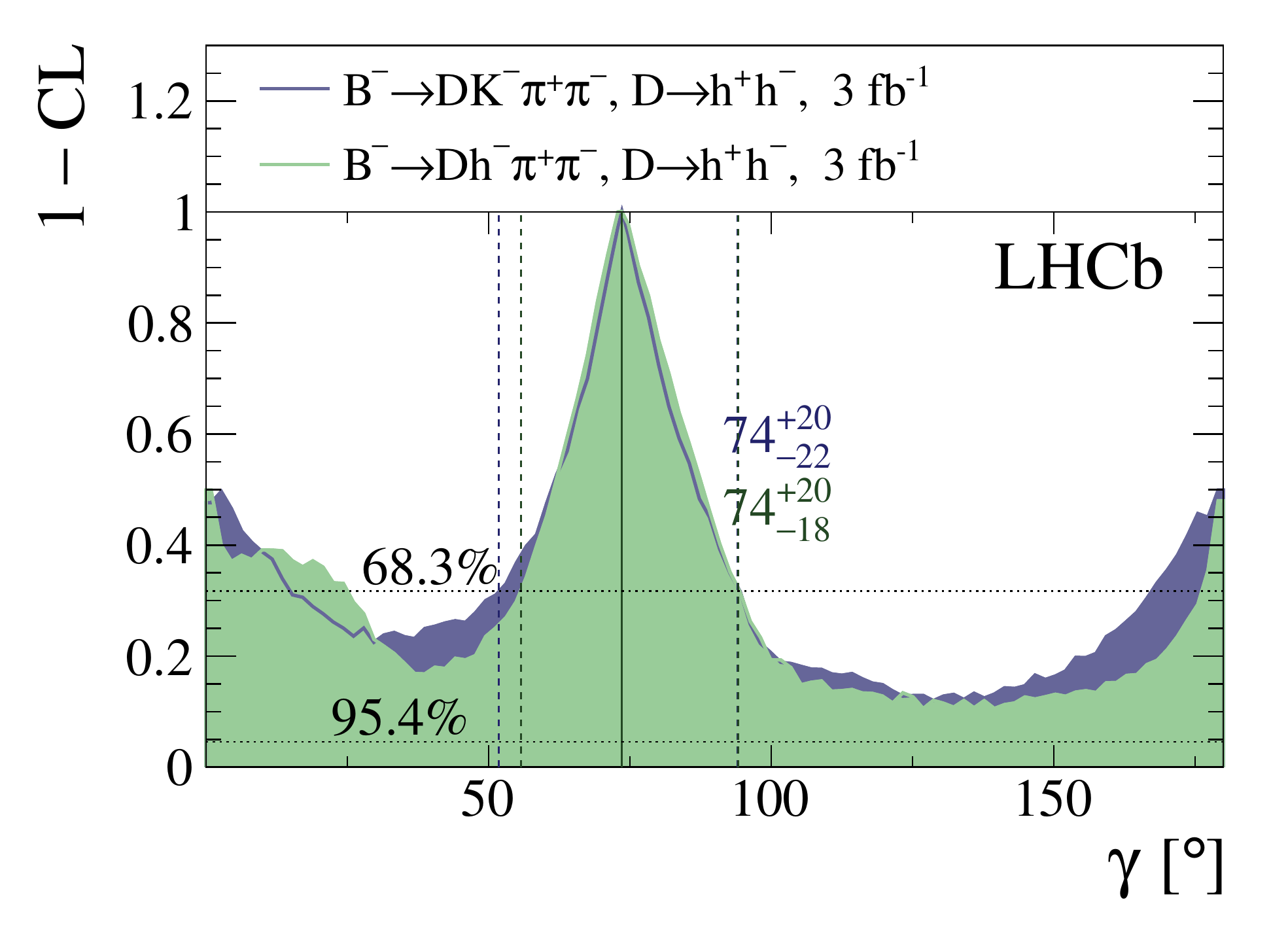}
  
  \caption{\label{fig:D3h}Left: The invariant mass distribution of the candidate decays of the type,
$B^+ \to D^0 K^+\pi^+\pi^-$ with $D^0 \to K^+\pi^-$. 
Right: The $1-\mathrm{CL}$ as a function of $\gamma$ with the combination of $B^+ \to D^0 h^+\pi^+\pi^-$ decay modes~\cite{LHCB-PAPER-2015-020}.}
\end{figure}

The least precisely measured of the internal angles of the unitarity triangle is
$\gamma = \arg(-V_{ud}V^*_{ub}/V_{cb}V_{cd}^*)$.
It is primarily measured using decays of the type $B \to D K$ in which there are amplitudes
with $b \to u$ and $b \to c$ quark level transitions that interfere with one another.
In addition to modes with a single kaon, it is possible to exploit modes in
which the $D$ recoils against other strange systems, for example, $K^+\pi^+\pi^-$~\cite{Gronau2003198}.
LHCb reports a first measurement of many different decay modes of this type using
the full Run-I dataset~\cite{LHCB-PAPER-2015-020}.
This includes the quasi flavour specific, or ADS~\cite{Atwood:1996ci,Atwood:2000ck} modes $D^0 \to K^-\pi^+, K^+\pi^-$, and the GLW modes $D^0 \to K^-K^+, \pi^-\pi^+$~\cite{Gronau1991172,GRONAU1991483}.
They are combined with $h^+\pi^+\pi^-$ ($h = K$~or~$\pi$).
Of particular note is the first evidence for the ADS decay $B^+ \to D^0 K^+\pi^+\pi^-$ with
$D^0 \to K^+\pi^-$, which has a high intrinsic sensitivity to $\gamma$.
Figure~\ref{fig:D3h} (left) shows the invariant mass spectrum of the positively charged $B^+$
candidates of this decay mode.
Figure~\ref{fig:D3h} (right) shows the determination of $\gamma$ from the combination of these $B^+ \to D^0 h^+\pi^+\pi^-$ decay modes.

\section{Measurement of \boldmath{$|V_{ub}|/|V_{cb}|$} with \boldmath{$\Lambda_b$} decays}
\label{Sec:Vub}

\begin{figure}[tb]\centering
  \includegraphics[width=0.49\textwidth]{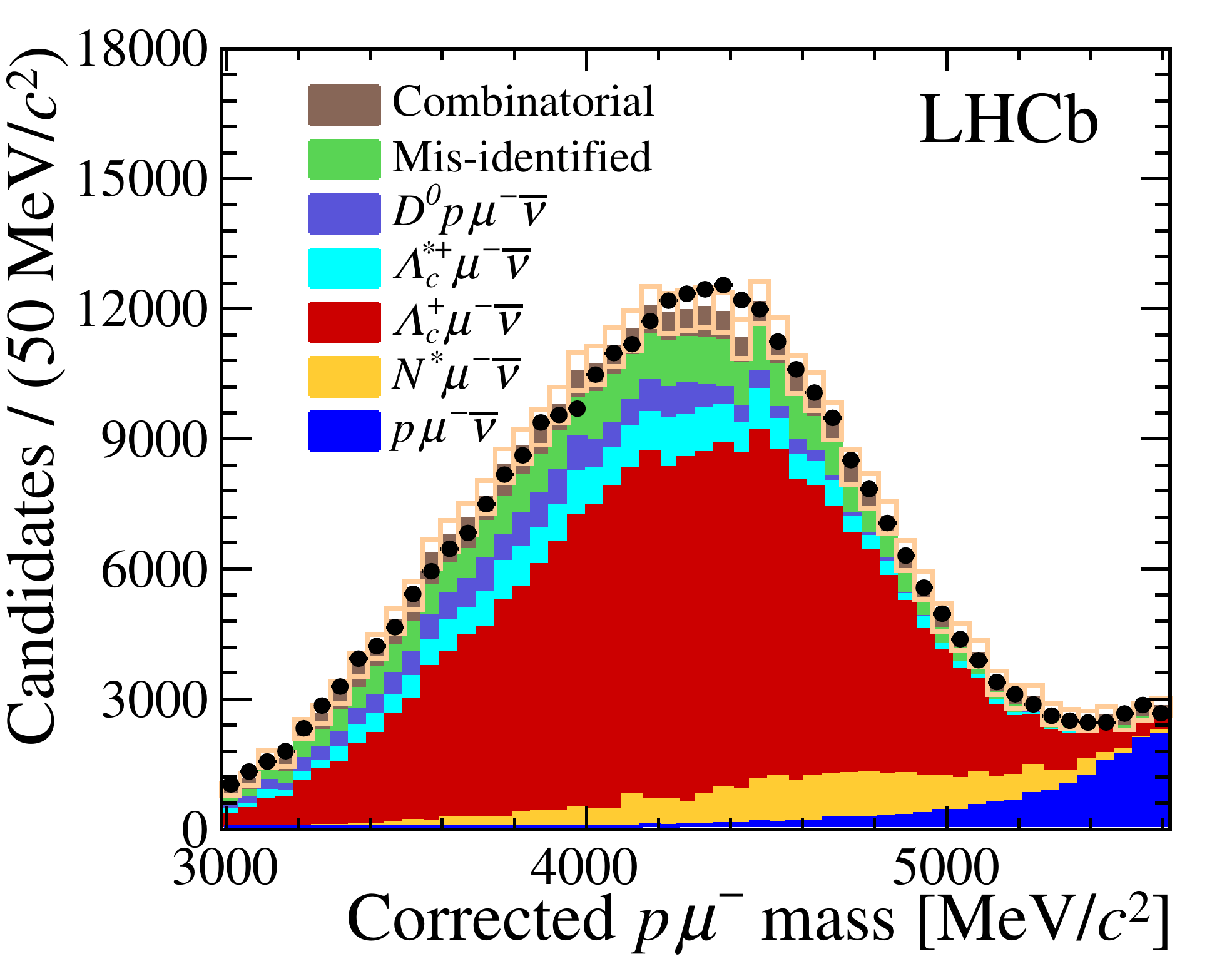}
  \includegraphics[width=0.49\textwidth]{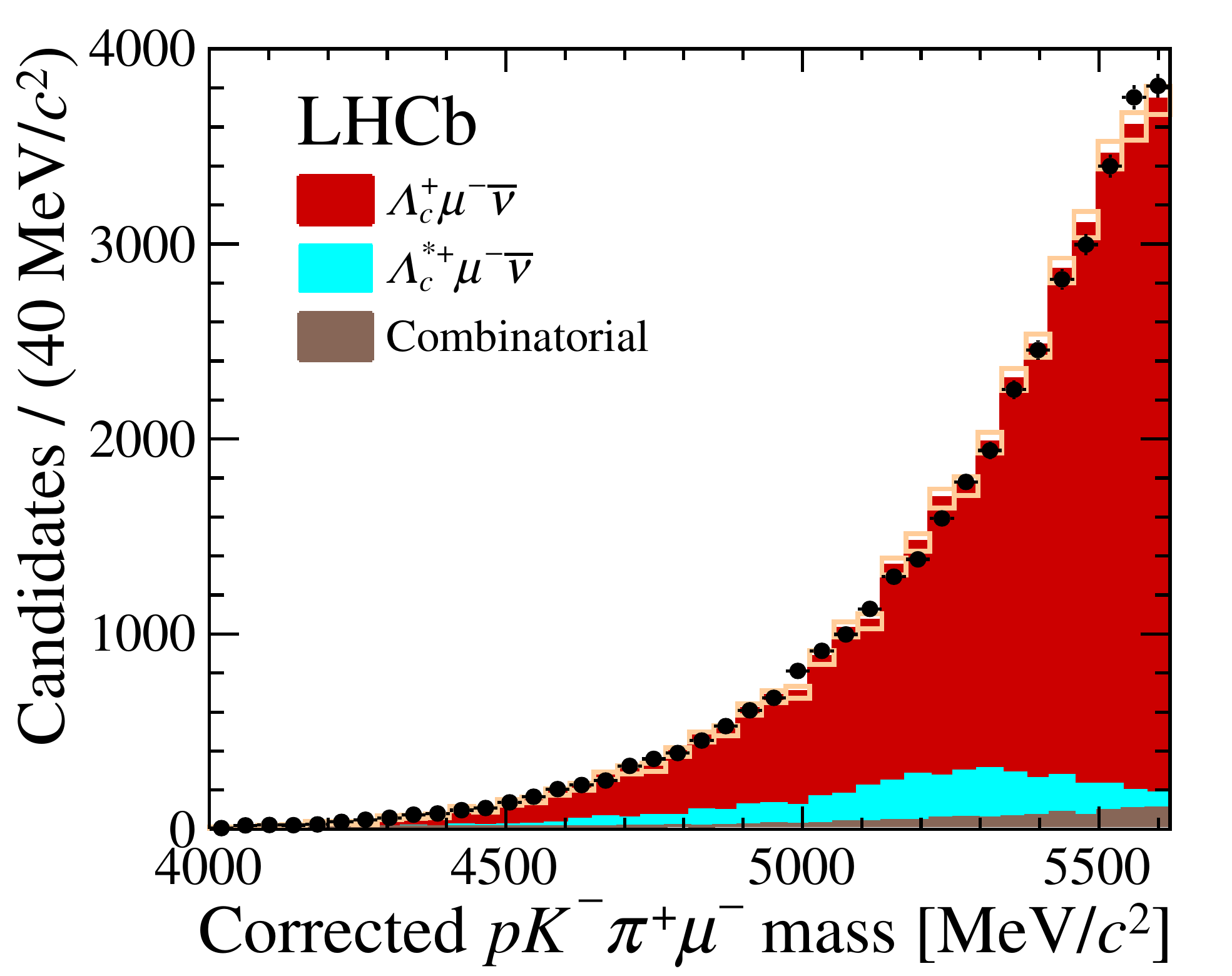}
  
  \caption{\label{fig:Vub}Left: the $m_{\rm corr}$ distribution
of $\Lambda_b \to p\mu\nu$ candidates. Right: 
the $m_{\rm corr}$ distribution of $\Lambda_b \to \Lambda_c\mu\nu$ candidates,
with $\Lambda_c \to pK\pi$.
}
\end{figure}

The magnitude of the CKM matrix element $V_{ub}$ is of great interest since it constrains
the length of the side of the unitarity triangle opposing the angle $\beta$.
The determination of $\sin 2\beta$ is through the loop induced mixing process which
is directly sensitive to new physics.
The value of $|V_{ub}|$ is extracted using semileptonic decays with transitions of the type
$b \to u \ell\nu$, primarily at the $e^+e^-$ B-factory experiments, BaBar and Belle~\cite{Lees:2012vv,delAmoSanchez:2010af,Ha:2010rf,Sibidanov:2013rkk}.
This is done using both inclusive measurements where only the charged lepton is reconstructed,
and with exclusive measurements of decays like $B^0 \to \pi^-\ell^+\nu$.
There is a long standing tension between determinations of $|V_{ub}|$ from these two types of measurements.
Before LHCb, the measurements of $|V_{ub}|$ were restricted to the $B^0$ and $B^+$ mesons.
In high energy proton-proton interactions, all species of $b$ hadrons can be produced,
with $\Lambda_b^0$ baryons of particular abundance.
In~\cite{LHCB-PAPER-2015-020} LHCb reports a first observation of the $V_{ub}$ decay $\Lambda_b^0 \to p \mu\nu$.
This measurement exploits the excellent vertexing capabilities of the LHCb detector
combined with the long lifetime of the $\Lambda_b$ through the corrected mass variable.
A line-of-flight between the primary $pp$ interaction vertex and the secondary $p\mu$
vertex of the candidate decays is determined.
The corrected mass exploits the fact that the visible $p\mu$ momentum transverse to this
flight direction, $p_{\perp}$, should balance with that of the neutrino.
It is defined as,
\[m_{\rm corr} = \sqrt{m^2 + p_{\perp}^2} + p_{\perp}.\]
For a decay with a single missing particle of zero mass,
the $m_{\rm corr}$ distribution exhibits a sharp kinematic end point
at the mass of the decaying parent particle.
Figure~\ref{fig:Vub} (left) shows the corrected mass distribution of the candidate decays
in which the signal component can be seen to the far right of the distribution.
Its decay rate is measured with respect to its Cabibbo favoured counterpart $\Lambda_b^0 \to \Lambda_c \mu\nu$,
again using a fit to the corrected mass distribution as shown in Figure~\ref{fig:Vub} (right).
The following ratio of decay rates is determined,
\begin{equation}
\frac{\mathcal{B}(\Lambda_b^0 \to p \mu\nu)_{q^2 > 15 \mathrm{GeV}/c^2}}{\mathcal{B}(\Lambda_b^0 \to \Lambda_c \mu\nu)_{q^2 > 7 \mathrm{GeV}/c^2}} = (1.00 \pm 0.04 \pm 0.08) \times 10^{-2},
\end{equation}
where $q$ is the invariant mass of the muon and neutrino.
This is related to the ratio $|V_{ub}|/|V_{cb}|$ by form factors.
Using a recent calculation of these form factors with lattice QCD methods~\cite{Detmold:2015aaa},
the following value is obtained, $|V_{ub}|/|V_{cb}| = 0.083 \pm 0.004_{\rm exp} \pm 0.004_{\rm FF}$.
This has a comparable precision to the exclusive determination with $B^0 \to \pi^+\ell\nu$
and is closer in value to it than to the inclusive determination.

\section{Conclusions}
The quark flavour sector remains a promising area in which to search for the indirect effects
of physics beyond the Standard Model.
Of particular interest is the study of $CP$ violating phenomena in beauty particles,
and several new measurements from the LHCb experiment are reported.
A measurement of $\sin2\beta$ using $B_d^0 \to  J/\psi K_s^0$~\cite{LHCB-PAPER-2015-004}
is already of comparable precision to those of the dedicated $e^+e^-$ B-factory experiments
for which this was considered to be their ``golden'' channel.
A measurement of of the $B_d^0-\bar{B}_d^0$ mixing frequency, $\Delta M_d$, using $B_d^0 \to  D^{(*)}\mu\nu X$ decays
is the most precise to date~\cite{LHCb-CONF-2015-003}.
A study of the decay $B_s \to J/\psi K^*$~\cite{LHCb-PAPER-2015-034}, in combination with an earlier study of the decay
$B_d^0 \to J/\psi \pi^+\pi^-$~\cite{Aaij:2014vda}, tightly constrains the possible penguin pollution to the determination of the phase $\phi_s$ with the decay $B_s^0 \to J/\psi \phi$.
A first study of decays of the type $B^+ \to D h^+ h^- h^+$ promises to provide a useful
constraint on the determination of the CKM angle $\gamma$~\cite{LHCB-PAPER-2015-020}.
LHCb reports the first observation of the decay $\Lambda_b \to p\mu\nu$ from which a first determination
of $|V_{ub}|/|V_{cb}|$ is made with $b$-baryon decays~\cite{LHCB-PAPER-2015-020}.
This measurement will help to shed light on the tension between $|V_{ub}|$ determinations with inclusive and exclusive $B$ meson decay rates.

\nocite{*}
\bibliographystyle{unsrt}
\bibliography{sample}

\begin{thebibliography}{10}

\bibitem{LHCB-PAPER-2015-004}
R.~Aaij et~al., {\em Phys. Rev. Lett.}, 115:031601, 2015.

\bibitem{Aaij:2012ke}
R~Aaij et~al., {\em Phys. Lett.}, B721:24--31, 2013.

\bibitem{Gronau:1992ke}
M.~Gronau, A.~Nippe, and J.~Rosner, {\em Phys. Rev.}, D47:1988--1993, 1993.

\bibitem{Aaij:2015qla}
R.~Aaij et~al., {\em JHEP}, 04:024, 2015.

\bibitem{Aubert:2009aw}
B.~Aubert et~al., {\em Phys. Rev.}, D79:072009, 2009.

\bibitem{PhysRevLett.108.171802}
I.~Adachi et~al., {\em Phys. Rev. Lett.}, 108:171802, Apr 2012.

\bibitem{LHCb-PAPER-2015-034}
R.~Aaij et~al., {\em JHEP}, 11:082, 2015.

\bibitem{LHCb-CONF-2015-003}
{LHCb collaboration}, ({LHCb-CONF-2015-003}).

\bibitem{Aaij:2014vda}
R.~Aaij et~al., {\em Phys. Lett.}, B742:38--49, 2015.

\bibitem{Gronau2003198}
M.~Gronau, {\em Physics Letters B}, 557(3–4):198 -- 206, 2003.

\bibitem{LHCB-PAPER-2015-020}
R.~Aaij et~al., 2015.
\newblock {submitted to Phys. Rev. D.}

\bibitem{Atwood:1996ci}
D.~Atwood, I.~Dunietz, and A.~Soni, {\em Phys. Rev. Lett.}, 78:3257--3260,
  1997.

\bibitem{Atwood:2000ck}
D.~Atwood, I.~Dunietz, and A.~Soni, {\em Phys. Rev.}, D63:036005, 2001.

\bibitem{Gronau1991172}
M.~Gronau and D.~Wyler, {\em Physics Letters B}, 265(1–2):172 -- 176, 1991.

\bibitem{GRONAU1991483}
M.~Gronau and D.~London, {\em Physics Letters B}, 253(3):483 -- 488, 1991.

\bibitem{Lees:2012vv}
J.~P. Lees et~al., {\em Phys. Rev.}, D86:092004, 2012.

\bibitem{delAmoSanchez:2010af}
P.~del Amo~Sanchez et~al., {\em Phys. Rev.}, D83:032007, 2011.

\bibitem{Ha:2010rf}
H.~Ha et~al., {\em Phys. Rev.}, D83:071101, 2011.

\bibitem{Sibidanov:2013rkk}
A.~Sibidanov et~al., {\em Phys. Rev.}, D88(3):032005, 2013.

\bibitem{Detmold:2015aaa}
W.~Detmold, C.~Lehner, and S.~Meinel, {\em Phys. Rev.}, D92(3):034503, 2015.

\bibitem{LHCb-PAPER-2015-013}
R.~Aaij et~al., {\em Nature Physics}, 11:743, 2015.

\end{thebibliography}

\end{document}